\documentclass{aastex63}
\usepackage{amsmath,amsfonts,amsthm,amstext}
\usepackage{amssymb}

\received{July 20, 2020}
\revised{ }
\accepted{ }

\shorttitle{Multiple stars and cluster mass}
\shortauthors{Borodina et al.}

\begin{document}
\title{Unresolved multiple stars and Galactic clusters' mass estimates}

\correspondingauthor{Giovanni Carraro}
\email{giovanni.carraro@unipd.it}

\author{Olga I. Borodina}
\affil{Institute of Astronomy  of the Russian Academy of Sciences \\
119017, 48 Pyatnitskaya st. \\
Moscow, Russia}
\affil{Moscow Institute of Physics and Technology \\
141701, 9 Institutskiy per., Dolgoprudny, \\
Moscow Region, Russia}

\author{Giovanni Carraro}
\affil{Dipartimento di Fisica e Astronomia, Universita'  di Padova\\
Vicolo Osservatorio 3\\
I35122, Padova, Italy}

\author{Anton F. Seleznev}
\affil{Ural Federal University \\
620002, 19 Mira street, \\
Ekaterinburg, Russia}

\author{Vladimir M. Danilov}
\affil{Ural Federal University \\
620002, 19 Mira street, \\
Ekaterinburg, Russia}

\begin{abstract}
If not properly accounted for, unresolved binary stars can induce a bias in the photometric determination of star cluster masses inferred from star counts and the luminosity function. A correction factor close to 1.15 (for a binary fraction of 0.35) was found in \citep{Boro19}, which needs to be applied to blind photometric mass estimates. This value for the correction factor was found to be smaller than literature values.
In an attempt to lift this discrepancy, in this work the focus is on higher order multiple stars with the goal of investigating the effect of triple and quadruple systems adopting the same methodology and data-set as in the quoted work. Then the result is found that  when triple  and quadruple together with binary systems are properly accounted for, the actual cluster mass (computed as all stars were single) should be incremented by a factor of  1.18$-$1.27, depending on the cluster and when the binary fraction $\alpha$ is 0.35. Fitting formulae are provided to derive the increment factor for different binary star percentages.
\end{abstract}


\section{Introduction}
Binary and higher order multiple stars play a very important role in the dynamical evolution of star clusters, mainly because
dynamically active binary and multiple systems can absorb (negative) energy.
In the case of globular clusters, such systems are known to prevent the so-called gravo-thermal catastrophe \citep{Heggie,Sugimoto}.
As for open clusters, dynamically active pairs can inflate the parent cluster.
According to theory,
early on, during a star cluster assembly, wide and close binary and multiple systems are forming, before the non-stationary, dynamically young star cluster, starts to contract. 
Then wide systems (and dynamically active pairs) tend to be disrupted during this contraction phase \citep{P1,P2,P3}.
Wide systems are forming again during the expansion phase \citep{Danilov2020} which naturally follows the contraction.
On the observational side,
a deficit of dynamically active binaries has been identified in some nearest open clusters \citep{DK,Danilov2020}.
Therefore, the evolution of wide binaries (their number and location inside the cluster) and multiple stars can serve as probe of open clusters' dynamical state \citep{Danilov2020}.
 
The binary fraction changes (diminishes) during the evolution of a star cluster.
Nevertheless, the distribution of orbital parameters and the distribution of the stellar mass ratio are conserved keeping a memory of the primordial binaries' properties (see references in \citet{Boro19}).

The dynamical evolution of stars in close binary systems in star clusters generate different types of exotic stellar objects, such as blue 
straggler stars \citep{Arp2,EBSS}, millisecond pulsars \citep{msp}, cataclysmic variables \citep{CV}, X-ray binaries \citep{Xray}, binary black holes \citep{BBH}, and so on.
The loss of material enriched by elements produced in CNO cycle during the mass transfer in massive interacting binaries constitutes one of possible scenarios for the origin of the multiple stellar populations in globular clusters \citep{Renzini+2015}.
 
The presence of unresolved binary and multiple systems can distort the estimates of the cluster mass determined both by the velocity dispersion \citep{4337,raste} and by the cluster luminosity function obtained from star counts (see a discussion in \citet{Boro19}).   

The present study follows up \citet{Boro19} study where the  influence of unresolved binary stars (UBS) on the estimate of open star cluster (OSC) mass derived through its luminosity function (LF) was investigated.
In \citet{Boro19} two general parameters for characterizing the binary star population --- the binary fraction $\alpha$ and the stellar mass ratio $q$ distribution  --- were used (see the review in \citet{Boro19}).
The findings were compared against the results obtained by \citet{KB} for Praesepe. for which a mass increment value of 1.35 for a binary fraction of 35$\%$.  Instead of 1.35, \citet{Boro19} found increment values in the range 1.06-1.19, and the range depends on the adopted $q$ distribution function.

In an attempt to lift this significant discrepancy, we explore in this study a possible improvement, namely we add multiple systems with higher multiplicity, namely triple and quadruple systems, to the UBS.

As a result, our paper is organised as follows.
Section 2 presents a survey of recent literature on multiple systems,  which our study stems from.
Section 3 is devoted to the description of our method.
Section 4 describes and discusses our results.
Section 5, finally, is dedicated to a summary of our findings.

\section{High order stellar systems in star clusters}

No much data are available in the literature on the presence and abundance ratios of multiple systems in star clusters.
For the multiple systems ratio we take the data from \citet{Mermilliod_Pleiades} for the Pleiades cluster, which should be more appropriate for the case of OSCs than the field stars' data from \citet{Tokovinin}.
Anyway, {\citet{Tokovinin} found for stellar systems with multiplicity of 1:2:3:4:5 (``1'' stands for single stars, ``2'' for binaries, ``3'' for triples, ``4'' for quadruples, and ``5'' for quintuples) the relative abundance ratios of 54:33:8:4:1.
These estimates were obtained compiling data in the  solar neighborhood.
The population of multiple stars in stellar clusters might be significantly different, though.
For instance, \citet{Mermilliod_Pleiades} found the relative abundance ratios of 56:30:2 for stars in the Pleiades cluster (singles, binaries, and triples).
A similar work was performed by \citet{Mermilliod_Praesepe} for Praesepe cluster as well.
They found the relative abundance ratios of 47:30:3 for stars of different multiplicity.
\citet{Mermilliod} considered different sources of incompleteness in the search of binary and multiple systems and concluded that ``in spite of the large efforts undertaken, the available material is still incomplete at several levels''.
The situation has not changed substantially since that time.
In addition, \citet{Mermilliod_Pleiades} data on Pleiades bases on 88 stars of F5-K0 spectral classes ($(B-V)\in[0.38;1.1]$) from the circle of about 70 arcminutes radius around the star Alcyone.
This field lies well inside the cluster core (the Pleiades core radius is of 2.62$^\circ$ and corona radius is of 10.9$^\circ$ according to \citet{DS_Pleiades}).
Then, the data of \citet{Mermilliod_Pleiades} are incomplete because they refer to the inner cluster area only.

\citet{Bouvier+_Pleiades} observed 144 G and K dwarf members of the Pleiades and found 22 binary systems and 3 triples.
\citet{Tokovinin+2006} found that most low-period spectroscopic binaries have a tertiary companion (at least for field stars).
Numerous works devoted to search of spectroscopic binaries or binaries with photometric data in nearby open star clusters can easily miss a tertiary companions if these companions are visually separated.

On the other hand, more recently, \citet{Danilov2020} considered a sample of 395 stars (probable members of the Pleiades) in an area of 2.5$^\circ$ around the cluster center with $G<15^m$ and errors in the tangential velocities less than 0.177 km s$^{-1}$.
36-37 wide visual pairs of single stars and 62-70 unresolved binary (or multiple) stars were extracted basing on their position in the color-magnitude diagram.
The distances between the components in visual pairs were found to be larger than 0.165 pc (approximately 4000 astronomical units).
The mean ratio of the component masses in the visual pairs is $q=0.67\pm0.04$, the $q$ distribution is approximately flat for $q\in[0.05;0.8]$ with the local maximum at $q=0.85$.
\citet{Danilov2020} marked 9 coincidences of the unresolved multiple stars with the components of the visual pairs, that is, possibly, triple or quadruple systems.
Moreover, he selected two triple, one quadruple and one sextuple visual systems with the relative velocities in pairs close to circular ones.
If we consider all unresolved multiples as binaries, then a grnnad-total of 260-270 singles, 89-98 binaries and 9 triples is found.

Nevertheless, for our investigation we take data of \citet{Mermilliod_Pleiades} as representative for our program clusters (they give minimum triples content among all multiples) and data of \citet{Tokovinin} for field stars in the assumption that for different clusters we should get some intermediate result for the mass increment between these two extreme cases. }

For the $q$ distribution we limit ourselves to the flat distribution given the recent findings of \citet{LiLu}.
We make use of  the same data on luminosity functions for clusters IC 2714, NGC 1912, NGC 2099, NGC 6834, and NGC 7142 as in \citet{Boro19}.
These LFs were obtained with the use of 2MASS data \citep{2MASS} by the statistical method described in \citet{LF,Pal1,kernel,4337,Rup147}.

\section{Description of the method}

\noindent
To determine the cluster mass increment produced by the additional mass budget stored in unresolved stars we simulate open clusters by creating stars according to the real luminosity function with binary fraction $\alpha$ defined as
$$\alpha = \frac{N_{binaries}+N_{triples} + N_{quadruples}}{N_{singles} + N_{binaries}+N_{triples}+N_{quadruples}}$$
triples fraction among multiples $\beta$
$$\beta = \frac{N_{triples} }{N_{binaries}+N_{triples}+ N_{quadruples}}$$
and, likewise, quadruples fraction $\gamma$
$$\gamma = \frac{N_{quadruples} }{N_{binaries}+N_{triples}+ N_{quadruples}}$$
and components' mass ratio distribution $f(q)$.

In our algorithm the binary fraction $\alpha$ varies between 0.1 and 0.9 in increments of 0.1.
Triples fraction among multiples $\beta$ is calculated from either \citet{Mermilliod_Pleiades} or \citet{Tokovinin} studies and is equal to 2:32 (it means that every 2 stars among 32 multiples are triples) or 8:45, respectively.
Quadruples fraction among multiples $\gamma$ for \citet{Mermilliod_Pleiades} case is equal to zero and for \citet{Tokovinin} ratio is equal to 4:45.
We do not take into account the number of quintuple systems and consider it negligible.
Then, the mass ratios $q_i = M_i/M_1$ in our simulations are uniformly distributed between 0 and 1  (i = 2 for binary, i = 2, 3 for triple star, and i = 2, 3, and 4 for quadruple star).

The magnitude distribution is binned in equal intervals $\Delta J$, and in each of them, we count the number of stars $N$ in accordance with the cluster' LF (analogous to \citet{Boro19}.
Then, by considering $\alpha$, $\beta$, \textbf{$\gamma$}, we calculate the number of binary, triple, and quadruple stars in each interval.

\begin{equation}
    N_{triples} = \alpha * \beta * N
\end{equation}

\begin{equation}
N_{quadruples} = \alpha * \gamma * N
\end{equation}

\begin{equation}
    N_{binaries} = \alpha * N - N_{triples} - N_{quadruples}
\end{equation}

\begin{equation}
    N_{singles} = N - N_{binaries} - N_{triples} - N_{quadruples}
\end{equation}

\noindent
Since all $N_{quadruples}, N_{triples}, N_{binaries}, N_{singles}$ should be integers, we round up numbers of multiple stars.
For all stars in the bin we use the same mean magnitude. 

Then we derive the luminosity corresponding to each mean magnitude.
For MS stars we calculate it using the \citet{Eker} formula extracting  mass from isochrone tables.
For stars that have already left MS we use the younger isochrone with the age of $4\cdot10^7$ years to determine the luminosity of the evolved stars at the MS stage with the same mass as evolved star mass. 

We then generate the component mass ratios $q_2, q_3, q_4$ for every quadruple star, $q_2, q_3$ for every triple star, and  $q_2$ for every binary star using the Neumann method \citep{Boro19}.
Therefore, for each multiple we have the following system of equations \citep{Danilov2020}:

\begin{equation}
\left\{
\begin{aligned}
&L= \sum^k_{i=1} L_i\\
&\log{L_i}= -0.705(\log{M_i})^2 + 4.655(\log {M_i}) - 0.025 \\
&q_i=\frac{M_i}{M_1} \\
&i = 1, 2, ...\: k\\
\end{aligned}
\right.
\label{MAINsystem}
\end{equation}

\noindent
where $q_1 \equiv 1$, $L$ and $L_i$ are the luminosity of the whole system and, separately, for each components, and $k$ is the components number in multiple stars.
$M_1, M_2, M_3, M_4$ are the dependent variables we are searching a value for, and $L_1, L_2, L_3, L_4$ are unknown quantities as well.
The second equation in this system is the mass-luminosity relation from \citet{Eker}.\\

\noindent
We can re-arrange this system into one final equation $f(x) = 0$:

\begin{equation}
    f(x) = dx^2 + bx + c + \log(1 + F(x)) - \log L \, ,
\end{equation}

\begin{equation}
    F(x) = \sum_{i=2}^k 10^{d (\log q_i)^2 + \log q_i (2dx + b)} \, ,
\end{equation}

\noindent
where $x = \log M_1$, $d = - 0.705, b = 4.655, c = - 0.025$\\

\noindent
Solving for $x$ we are able to derive all components masses $M_1 = 10^x$,  $M_i = q_i M_1$ ($i = 2$ for binary, $i = 2, 3$ for triple star, and  $i = 2, 3, 4$ for quadruple star).\\

As a result, we can add up the masses of multiples with the single star masses and eventually obtain an estimate of the total mass of the cluster.

Because mass ratios $q_i$ are generated randomly, the resulting mass can vary, so we repeat all procedures several times (30) and calculate the mean and standard deviation of the derived cluster mass.

If we consider all stars as singles, we can calculate a different estimate of cluster mass $M_{wm}$ using the isochrone table and therefore the mass increment $y=M/M_{wm}$ (\textit{wm} stands here for ``without multiples'').

The code is available online at the \href{https://github.com/olgaborodina/Unresolved_stars_in_clusters}{link} 
(https://github.com/olgaborodina/Unresolved\_stars\_in\_clusters).

\section{Results and discussion}

\begin{figure}
    \centering
    \includegraphics[width=19cm]{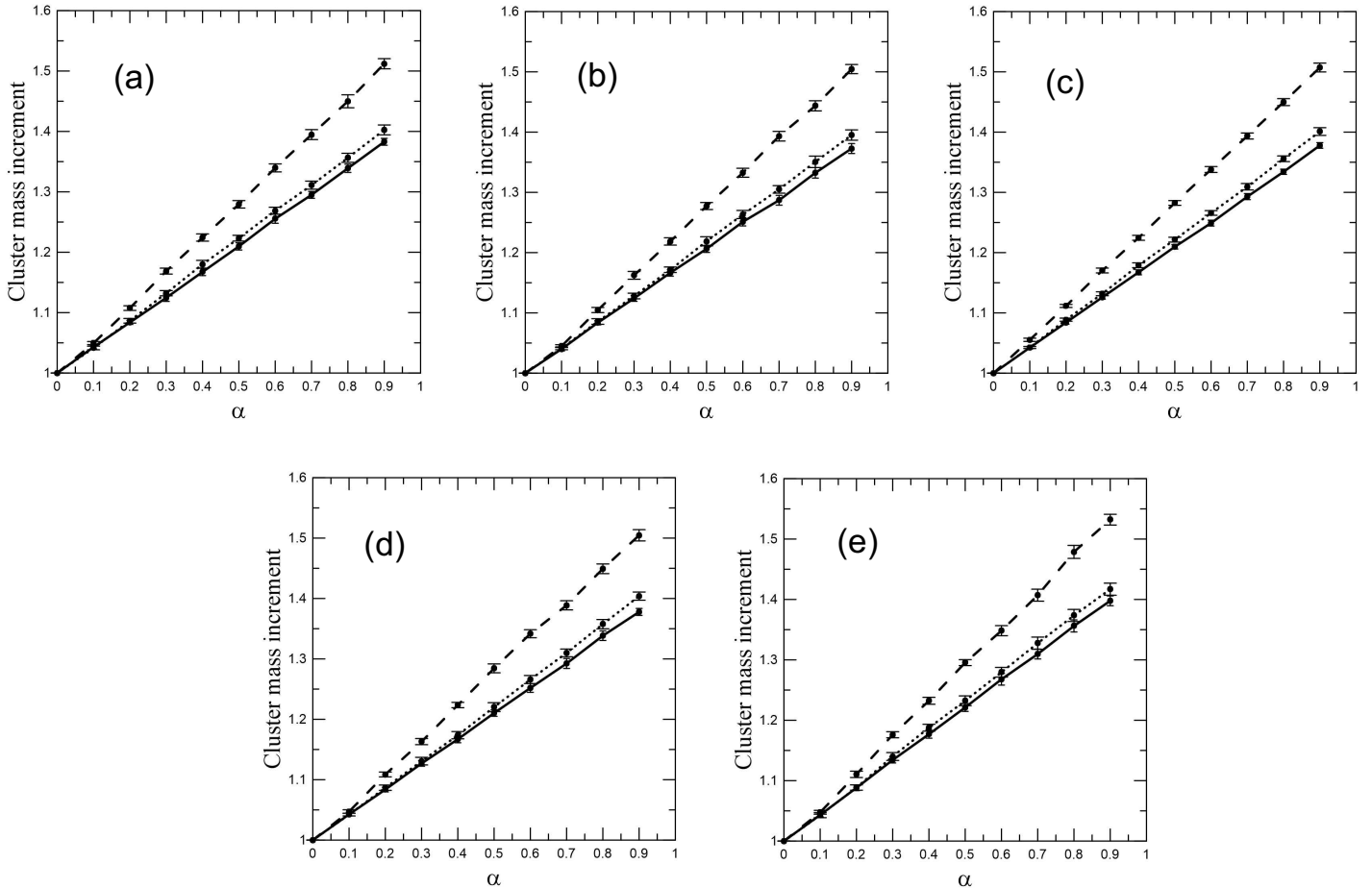}
    \caption{The dependence of the cluster mass increment on the binary star fraction $\alpha$. Solid line: binary systems only; dotted line: binary and triple systems according to \citet{Mermilliod_Pleiades}; dashed line: binary,  triple, and quadruple systems according to \citet{Tokovinin}. (a) IC 2714; (b) NGC 1912; (c) NGC 2099; (d) NGC 6834; (e) NGC 7142.}
     \label{increments-alpha}
\end{figure}

Figure 1 shows the dependence of the cluster mass increment on the binary star fraction $\alpha$ for three cases.
The first case (the lower solid line) corresponds to clusters with  unresolved binaries only (this result comes from our previous paper \citet{Boro19}).
The second case (the middle dotted line) corresponds to clusters with binary and triple systems adopting their ratio from \citet{Mermilliod_Pleiades}.
Finally, the third case (the upper dashed line) corresponds to  clusters with binary, triple, and quadruple systems adopting their ratio from \citet{Tokovinin}.
In all cases we use a flat distribution for the multiple star component mass ratio.\\

\noindent
The case with the multiple system ratio of \citet{Mermilliod_Pleiades} differs slightly from the case of binary systems only, unlike to the case of multiple system ratio of \citet{Tokovinin}.
We can explain it because in the case of \citet{Mermilliod_Pleiades} the ratio of triple systems to all multiples is $1/16\approx0.06$ only while in the case of \citet{Tokovinin} the ratio of triples and quadruples makes up $12/33\approx0.36$ of all multiples.

It is readily seen that even in the case of adopting the multiple star ratio for Galactic field \citep{Tokovinin} the cluster mass increment does not exceed 1.20 for the specific value of the binary fraction of 0.35.
In the more realistic case of multiple star ratio for the Pleiades cluster \citep{Mermilliod_Pleiades}, the cluster mass increment does not exceed 1.16.
Then, the value of \citet{KB} of 1.35 for the mass increment can not be explained with the luminosity-limited pairing \citep{Boro19}.

We deem that the likely explanation of the \citet{KB} value for the cluster mass increment could be following.
If the binary fraction is 0.35 and the second components of the binary systems are distributed with the same mass function as the primary components, then an addition to the cluster mass should be just 0.35 of the cluster mass in the case of single stars only (which is ``primary-constrained random pairing'' as described by \citet{Kouwenhoven}).
Such approach is quite reasonable, for example, when one sets up an initial cluster model for N-body experiments.
However, such arguments contain a mistake when one estimates the cluster mass from photometric data because the luminosity of the binary star composed in this way would be larger than the observed one.

\citet{KB} determined the mass of the Praesepe cluster following this logical path.
First, they selected probable cluster members and evaluated their masses using isochrone tables and treating them as  single ones.
Second, \citet{KB} evaluated the mass of invisible low-mass stars and stellar remnants of massive stars.
Then, they assumed that 35 percent of stars were binaries and added the mass of secondary components taken from the same mass function of single stars.
As a result, the mass of each binary star would naturally increase by 35 percent on average.
In turn, however, the luminosity of every binary star would also increase (and its stellar magnitude would decrease).
The correct  way should probably be to take different (smaller) mass values both for primary and secondary components,  using for instance the ``luminosity limited pairing'' as in \citet{Boro19} or in the present paper.

\begin{deluxetable}{lcccccc}
\tablecaption{Linear approximation $y=A+B\alpha$ for the cluster mass increment dependence on the multiple fraction \label{tab:approx}}
\tablehead{
\colhead{Multiple star ratio} & \colhead{Cluster} & \colhead{A}  & \colhead{B} & \colhead{$\chi^2$} & \colhead{Q} & \colhead{$y(0.35\pm0.05)$}
}
\colnumbers
\startdata
The ratio of                &  IC 2714 & 0.997 $\pm$ 0.003 & 0.451 $\pm$ 0.007 & 0.503  &  1.000 & 1.15$\pm$0.02 \\
multiple stars              & NGC 1912 & 0.997 $\pm$ 0.002 & 0.441 $\pm$ 0.007 & 0.465  &  1.000 & 1.15$\pm$0.02 \\
for Pleiades                & NGC 2099 & 0.998 $\pm$ 0.002 & 0.447 $\pm$ 0.004 & 0.667  &  1.000 & 1.15$\pm$0.02 \\
\citet{Mermilliod_Pleiades} & NGC 6834 & 0.997 $\pm$ 0.002 & 0.449 $\pm$ 0.006 & 0.656  &  1.000 & 1.15$\pm$0.02 \\
                            & NGC 7142 & 0.996 $\pm$ 0.003 & 0.471 $\pm$ 0.009 & 0.869  &  0.999 & 1.16$\pm$0.02 \\
\hline
The ratio of                &  IC 2714 & 0.993 $\pm$ 0.003 & 0.576 $\pm$ 0.007 & 1.333  &  0.995 & 1.19$\pm$0.03 \\
multiple stars              & NGC 1912 & 0.989 $\pm$ 0.002 & 0.574 $\pm$ 0.006 & 2.161  &  0.976 & 1.19$\pm$0.03 \\
for Galactic field          & NGC 2099 & 0.999 $\pm$ 0.002 & 0.564 $\pm$ 0.005 & 0.407  &  1.000 & 1.20$\pm$0.03 \\
\citet{Tokovinin}           & NGC 6834 & 0.992 $\pm$ 0.003 & 0.574 $\pm$ 0.007 & 3.511  &  0.898 & 1.19$\pm$0.03 \\
                            & NGC 7142 & 0.990 $\pm$ 0.003 & 0.606 $\pm$ 0.008 & 3.967  &  0.860 & 1.20$\pm$0.03 \\
\hline
Binary stars only           &  IC 2714 & 1.003 $\pm$ 0.003 & 0.424 $\pm$ 0.006 & 0.612  &  1.000 & 1.15$\pm$0.02 \\
\citet{Boro19}              & NGC 1912 & 0.999 $\pm$ 0.003 & 0.415 $\pm$ 0.007 & 0.574  &  1.000 & 1.14$\pm$0.02 \\
                            & NGC 2099 & 1.000 $\pm$ 0.002 & 0.418 $\pm$ 0.004 & 0.314  &  1.000 & 1.15$\pm$0.02 \\
                            & NGC 6834 & 1.000 $\pm$ 0.002 & 0.419 $\pm$ 0.006 & 0.234  &  1.000 & 1.15$\pm$0.02 \\
                            & NGC 7142 & 0.999 $\pm$ 0.003 & 0.444 $\pm$ 0.008 & 0.230  &  1.000 & 1.15$\pm$0.02 \\
\enddata
\end{deluxetable}

Table 1 lists the coefficients of a linear regression for the dependencies shown in Figure 1 and the values of the mass increment for a representative binary ratio of 0.35$\pm$0.05 \citep{KB}, for the sake of illustration and comparison with \citet{KB}.
The data for the case of binary stars only is from \citet{Boro19} (we referred there to it as ``flat distribution'').
The large values of $\chi^2$ are explained by the small values of the standard deviation for the cluster mass when repeating the random pairing procedure.
If we artificially increase ten times the standard deviations for the mass estimates of NGC 1912 (second line in Table 1), the $\chi^2$ value becomes 0.367 and the Q value becomes 1.0.\\

\noindent
A general, important, remark is now in order.
When calculating the cluster mass increment due to the presence of unresolved  binary and multiple systems, one needs in principle to take into account the spatial resolution of the data employed to construct thea LF.
In the present work we use the cluster LFs obtained counting stars  extracted from the 2MASS Point Source catalog \citep{2MASS}.
The spatial resolution of 2MASS is of about $\delta=4$ arcseconds (https://old.ipac.caltech.edu/2mass/overview/about2mass.html).
It corresponds to a separation between the binary star components of 4000 astronomical units (AU) at a cluster distance of 1 kpc.
Then, even the very wide binaries and hierarchical triples in the clusters of our sample could be unresolved (see the sample cluster distances in Table 1 of \citet{Boro19}).

However, if we were to use Gaia data \citep{Gaia}, for example, the situation would change significantly.
The spatial resolution of  Gaia DR2 for binary components is of about $\delta=0.5-0.6$ arcseconds \citep{Ziegler+2018} (7-8 times smaller than the 2MASS resolution).
This corresponds to separations of about 500-600 AU for a cluster distance of 1 kpc.
In that case  wide binaries would be resolved and the cluster mass increment would in turn be smaller.
{The resolution of Gaia DR3 should be even better.
Gaia mission goal in binary resolution for the final data release is $\delta=0.1$ arcseconds (https://www.cosmos.esa.int/web/gaia/science-performance).  }

We investigated the distribution of the apparent on-sky separations of the binary components.
With this aim, we used the distribution of the logarithm of period $P$ and the distribution of eccentricities for the solar-type binaries as from \citet{D&M91}.
This distribution of the period logarithms is very close to normal distribution with $\overline{\log P}=4.8$ and $\sigma_{\log P}=2.3$ (it means that the period distribution is log-normal).
We randomly (uniformly) set the values of the orbit plane inclination, the peri-astron longitude, and the time after the peri-astron passing.
The semi-major axes $a$ were determined using the period value and Third Kepler Law supposing the mean mass of the primary component to be 1 $M_{\odot}$  and the mean mass of secondary component to be 0.5$M_{\odot}$.
This results in a normal distribution of the semi-major axis logarithms with $\overline{\log a}=1.55$ (it corresponds to $\overline{a}=35$ AU) and $\sigma_{\log a}=1.53$.
Adopting constant values for the component masses is a rough approximation, especially for nearby star clusters with large interval of masses of stars available from observation.
However, for our program clusters' sample this interval is not that wide and the approximation seems appropriate.

The distribution of the logarithm of the apparent separations  turned out to mirror the distribution of the semi-major axis logarithms (we used 10000 random pairs).
Therefore, we used a Gaussian function for the apparent separation distribution function with the same parameters as the semi-major axis distribution function.
In order to determine the unresolved binary fraction we need to integrate this distribution from $-\infty$ to the $\log a_0$, where $a_0$ is the value of separation corresponding to the resolution of the catalog for the binary components $a_0(AU)=r(pc)*\delta(arcsec)$.
The required unresolved binary fraction (UBF), that is the ratio of unresolved binaries among all binaries, is:

\begin{equation}
    UBF(\log a_0) = \frac{1}{\sigma_{\log a}\sqrt{2\pi}}\int\limits_{-\infty}^{\log a_0} \exp \left\{{-\frac{(\log a-\overline{\log a})^2}{2\sigma_{\log a}^2}}\right\}d\log a = \frac{1}{2} + \frac{1}{2}erf\left(\frac{\log a_0-\overline{\log a}}{\sigma_{\log a}\sqrt{2}} \right) \, ,
\end{equation}

\noindent
assuming that $\log a_0>\overline{\log a}$.
This looks reasonable because even for $r=100$ pc the resolution of binary components would be 50--60 AU (in the case of Gaia DR2) and $\overline{a}=35$ AU.

Table 2 lists the UBF values for star clusters considered in \citet{Boro19} and in this paper for two cases of spatial resolution for binary components.
The first one corresponds to the 2MASS resolution (4 arcsecond) and the second one corresponds to the Gaia DR2 resolution (we take 0.5 arcsecond as ambiguity limit).
The cluster distances were taken the same as in \citet{Boro19}.

\begin{deluxetable}{lccccc}
\tablecaption{The unresolved binary fractions for sample clusters \label{tab:UBF}}
\tablehead{
\colhead{} & \colhead{}  & \multicolumn{2}{c}{\hspace{1cm}2MASS resolution}  & \multicolumn{2}{c}{\hspace{1cm}Gaia DR2 resolution} \\
\colhead{Cluster} & \colhead{r}  & \colhead{\hspace{1cm}$a_0$} & \colhead{UBF} & \colhead{\hspace{1cm}$a_0$} & \colhead{UBF} \\
\colhead{} & \colhead{pc}  & \colhead{\hspace{1cm}AU}  & \colhead{}   & \colhead{\hspace{1cm}AU}  & \colhead{} \\
\colhead{(1)} & \colhead{(2)}  & \colhead{\hspace{1cm}(3)}  & \colhead{(4)}   & \colhead{\hspace{1cm}(5)}  & \colhead{(6)} 
}
\startdata
IC 2714  & 1250 & \hspace{1cm}5000 & 0.92  &  \hspace{1cm}625 & 0.80 \\
NGC 1912 & 1140 & \hspace{1cm}4560 & 0.92  &  \hspace{1cm}570 & 0.79 \\
NGC 2099 & 1410 & \hspace{1cm}5640 & 0.93  &  \hspace{1cm}705 & 0.80 \\
NGC 6834 & 2080 & \hspace{1cm}8320 & 0.94  & \hspace{1cm}1040 & 0.83 \\
NGC 7142 & 1780 & \hspace{1cm}7120 & 0.94  &  \hspace{1cm}890 & 0.82 \\
\enddata
\end{deluxetable}

One can readily see that even for Gaia DR2 resolution the fraction of unresolved binaries keeps high.
The probability to detect resolved binary depends on the component mass ratio and the limiting stellar magnitude of the sample.
Independently on the visibility of the secondary component, a  resolved binary would behave as a single star for the purpose of deriving  the cluster mass.

When using Gaia data, one would need to find which fraction of binary and multiple systems is unresolved in the cluster according to its distance.
After that, the mass increment could be evaluated using this paper results.

We plan to investigate the population of the resolved binaries in nearest open clusters in the future, especially when the new Gaia data release (DR3) is publicly available.
The key point for such investigation is the accuracy of astrometric parameters and the presence of  accurate radial velocities.
We also plan to investigate the population of unresolved binaries with photometric data and spectroscopic monitoring of bright stars in nearest clusters.

\section{Conclusions}

In this study we investigated the effect of unresolved multiple stars (binaries, triples, and quadruples altogether) on Galactic clusters' mass estimates as obtained from clusters' LF built through star counts.
We used the same LF data and the ``luminosity limited pairing'' method as described in \citet{Boro19}.\\

\noindent
The data on the multiple stars' ratio were taken from \citet{Mermilliod_Pleiades} for the Pleiades open cluster and from \citet{Tokovinin} for the general Galactic field in the solar vicinity.\\

\noindent
The inspection of Figure 1 and Table 1 allows one to conclude that mass estimates obtained considering all stars as single should be corrected for factors which depend on the ratio of binary and multiple stars.
The correction factor, which implies always a mass increment, ranges from 1.18 to 1.27 (for a binary ratio of 0.35 as \citet{KB} determined for the Praesepe cluster).\\

\noindent 
The correction factor depends on the considered cluster only marginally.
On the contrary it shows quite a significant variation whether either field stars or Pleiades multiple star percentages are adopted.

\noindent
As expected, increasing multiple stars ratio, the mass increment turns out to be larger.
Therefore, the mass correction is larger if one adopts field stars' percentages for binary and multiple systems.\\

\noindent
Ideally one should obtain an independent, cluster by cluster, binary and multiple star percentage.
In fact, the Pleiades cannot be fully representative of every star cluster, since any individual star cluster has different mass at birth and undergoes different dynamical evolutionary history.
All this affects the number and nature of binary and multiple systems present at any given time.\\

\noindent
It is expected that the  third data release (DR3) will be very helpful to obtain more information on binary percentages.
Key also is the cluster distance which determines the amount of binaries we detect as unresolved, given the Gaia DR3 fixed spatial resolution.

\acknowledgments
The work of A.F.Seleznev and V.M.Danilov was supported by the Ministry of Science and Higher Education of the Russian Federation, FEUZ-2020-0030, and by the Act No. 211 of the Government of the Russian Federation, agreement no. 02.A03.21.0006. 
The O.I. Borodina's work was partly supported by the grant 075-15-2020-780 for major research projects of the Ministry of education and science and by RFBR and DFG  according to the research project No. 20-52-12009.

This publication makes use of data products from the Two Micron All Sky Survey, which is a joint project of the University of Massachusetts and the Infrared Processing and Analysis Center/California Institute of Technology, funded by the National Aeronautics and Space Administration and the National Science Foundation.

The input of the anonymous referee has been greatly appreciated.

{}

\end{document}